\documentclass[aps,twocolumn,amsfonts,showpacs,superscriptaddress]{revtex4-1}
\usepackage{amssymb,amsmath,amsthm,booktabs,mathtools}
\usepackage{bm}% bold math
\usepackage{color}
\usepackage{tikz}
\usepackage{enumitem}
\usepackage{units}
\usepackage{braket}
\usepackage{gensymb}
\usepackage{amsmath}
\usepackage{cancel}
\usepackage{soul} %for "biffer" the text
\usepackage{units}

\usepackage[version=4]{mhchem}

\usepackage{braket}

\begin{document}

\title{ Magic-wavelength nanofiber-based \\ two-color dipole trap with sub-$\lambda/2$ spacing}

\author{Lucas Pache}
\author{Martin Cordier}
\author{Hector Letellier}
\author{Max Schemmer}
\altaffiliation{New address: Istituto Nazionale di Ottica, Consiglio Nazionale delle Ricerche, 50019 Sesto Fiorentino, Italy}
\author{Philipp Schneeweiss}
\author{Jürgen Volz}
\author{Arno Rauschenbeutel}
%\email{...}
\affiliation{Department of Physics, Humboldt-Universität zu of Berlin, 10099 Berlin, Germany}

\begin{abstract}
We report on the realization and characterization of a novel magic-wavelength nanofiber-based two-color optical dipole trap for cesium that allows us to generate two diametral periodic one-dimensional arrays of trapping sites with a spacing significantly smaller than half the resonant free-space wavelength of the cesium D2-transition. This is achieved by launching a blue-detuned partial standing wave and two red-detuned light fields through the nanofiber. We trap and optically interface the atoms in the resulting periodic optical potential and characterize the trap by measuring the lifetime of the trapped atoms, the atom-light coupling strength, the filling factor, and the trap frequencies in the radial and axial direction. The implementation of this nanofiber-based optical interface with magic trapping wavelengths and sub-$\lambda/2$ spacing is an important step towards the exploration of novel collective radiative effects, such as selective radiance.   

\end{abstract}

\maketitle
Nanofiber-based atom--light interfaces have become a powerful tool for controlling and harvesting the interaction between propagating light fields and laser trapped atoms~\cite{solano_chapter_2017-2, nayak_nanofiber_2018, gonzalez-tudela_lightmatter_2024, Li_2024}. They allow for coupling arrays of individually trapped atoms to the nanofiber-guided mode via the evanescent field surrounding the nanofiber. This experimental platform has enabled studies and applications ranging from the efficient resonant~\cite{vetsch_optical_2010,goban_demonstration_2012-2} and dispersive~\cite{dawkins_dispersive_2011,beguin_generation_2014} coupling of light and optically trapped atoms to the demonstration of propagation direction-dependent attenuation~\cite{sayrin_nanophotonic_2015} and amplification~\cite{pucher_atomic_2022} of light. 
In many of these works, the atoms could be regarded as independent emitters, e.g., contributing linearly to the total attenuation of the light field. More recently, considerable interest has been directed toward the study of collective radiative effects with nanofiber-coupled atoms mediated by the fiber-guided mode~\cite{Ruddell_2017, Kato_2019}. This lead to the realization of optical memories~\cite{sayrin_storage_2015,gouraud_demonstration_2015, corzo_waveguide-coupled_2019} and atom-based Bragg-mirrors~\cite{corzo_large_2016, sorensen_coherent_2016} with this platform as well as the observation of sub- and superradiant emission of light, both in the single excitation sector~\cite{solano_super-radiance_2017, pennetta_collective_2022, pennetta_observation_2022} and with fully inverted atomic ensembles~\cite{liedl_collective_2023}. Moreover, a collective enhancement of the nonlinear response of the atoms was demonstrated allowing one to generate and tailor strongly correlated quantum states of light~\cite{prasad_correlating_2020, hinney_unraveling_2021,cordier_tailoring_2023}. Even richer physics is expected when the interatomic distance is smaller than half the resonant emission wavelength of the atoms, $\lambda$. In this case, the atoms interact with each other via near-field dipole-dipole interactions as well as via the common nanofiber-guided mode. An emblematic effect predicted in this regime is selective radiance, where the scattering of guided light back into the waveguide mode is collectively enhanced while it is suppressed for all unguided modes. This would enable low-loss propagation of light past the atomic array and may allow one to realize quantum devices such as quantum memories with significantly enhanced performance~\cite{asenjo-garcia_exponential_2017-1, Ruostekoski_2017, asenjo-garcia_optical_2019,Cardonner_2019, Needham_2019, Sheremet_2019}. First investigations with ultra-cold atoms arranged in two-dimensional optical lattices successfully demonstrated the cooperative enhancement of the light–matter coupling strength and the strong reflection of the incoming light~\cite{rui_subradiant_2020}. Here, we demonstrate a nanofiber-based two-color dipole trap with distance $d \simeq 0.35 \lambda$ between adjacent trapping sites, i.e., significantly smaller than $\lambda / 2$. We use magic wavelengths for the trapping fields in order to minimize the inhomogeneous broadening of the optical transition frequency due to trap-induced Stark shifts, thereby providing optimal conditions for the implementation and observation of novel collective radiative effects. 

\begin{figure}[t]
\centering
\includegraphics[width=0.5\textwidth]{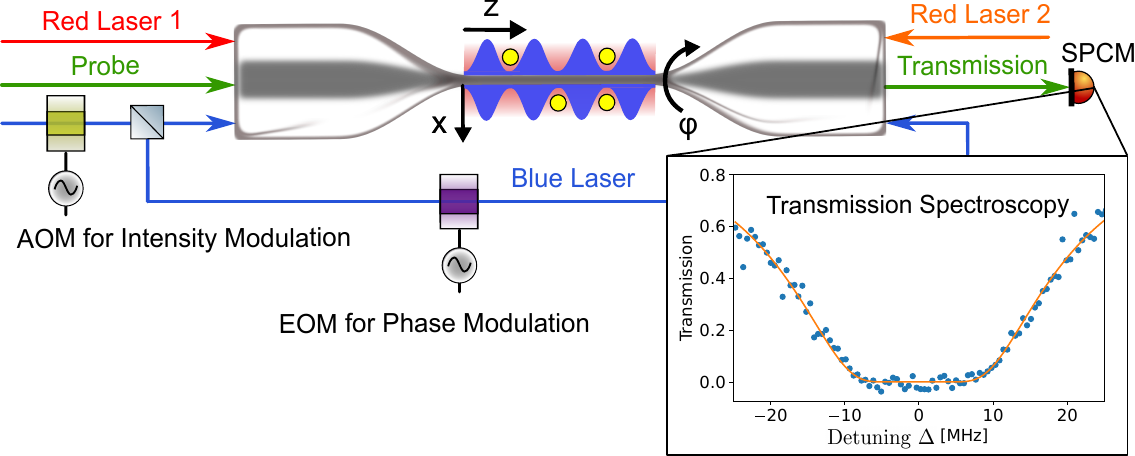}
\caption{Experimental setup of the nanofiber-based two-color dipole trap: The interference of two blue-detuned counter-propagating light fields creates a repulsive partial standing wave dipole potential along the $z$-direction. Together with two attractive red-detuned running wave dipole potentials, which are $ \unit[150]{GHz}$ detuned from each other in order to minimize the vector light shift, we create two diametral one-dimensional arrays of trapping sites in the evanescent field of the nanofiber. One acousto-optic modulator (AOM) in the path of the blue-detuned light is used for intensity modulation and an electro-optic modulator (EOM) is used for phase modulation of the high-power arm of the counter-propagating blue-detuned trapping fields. By means of these modulations, we can study the modulation frequency dependent heating of the atoms via loss spectroscopy by probing the atomic ensemble with a laser field resonant with the D2-line at a wavelength of $\unit[852]{nm}$. Example of a measurement of the optical depth after loading the dipole trap from the molasses. We scan the probe laser frequency over the $6S_{1/2} \; F=4 \rightarrow 6P_{3/2} \; F^{\prime}=5$ D2-transition with an intensity well below saturation. We repeat the measurement 50 times and deduce the on-resonance optical depth (OD) by fitting the data with a transmission function, $T(\Delta) = \exp \left(-OD/(1+4 \Delta^2/\Gamma^2 \right) $, where $\Delta$ denotes the atom-laser detuning and $\Gamma$ is the free-space emission rate. In this specific measurement, we find an on-resonance optical depth of OD = 44.0 $\pm$ 0.6. }
\label{fig:blue_trap}
\end{figure}

\begin{figure}[h]
\centering
\includegraphics[width=0.5\textwidth]{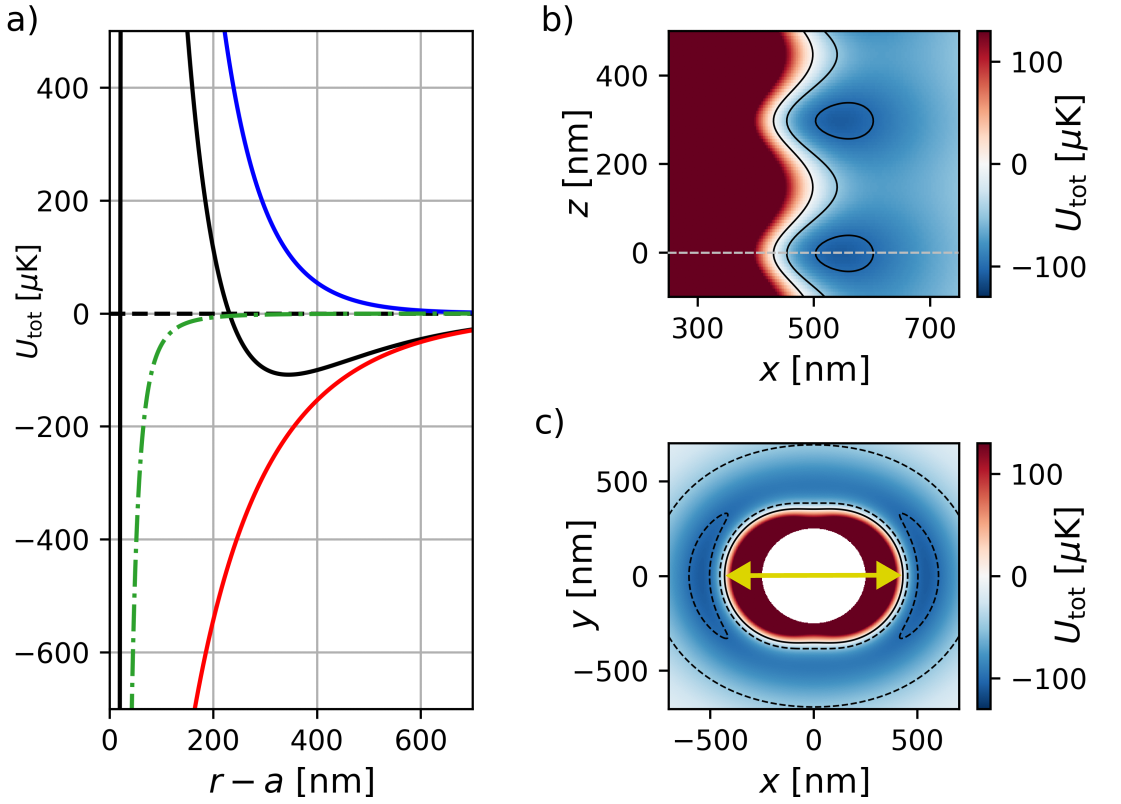}
\caption{ Calculated trapping potential: a) Radial dipole potential at a minimum of the blue-detuned partial standing wave potential, see dashed-white line in b). The total potential (black line) is the sum of the attractive potential created by the red-detuned trapping fields (red line), the repulsive potential created by the blue-detuned light fields (blue line) and the attractive van der Waals potential (green dashed-dotted line). b) Trapping potential in the $x$--$z$-plane, i.e., along the nanofiber, as a function of the radial distance from the nanofiber center. c) Trap potential in the $x$--$y$-plane, where the $z$-position corresponds to a minimum of the blue-detuned standing wave potential. The potential is calculated for powers of $\unit[0.6]{mW}$ and $ \unit[16]{mW}$ for the blue-detuned trapping fields and equal powers of $2\times 0.5$~mW for the red-detuned trapping fields. All trapping fields are quasi-linearly polarized along the $x$-axis, also indicated by the yellow arrow. }
\label{fig:blue_potential}
\end{figure}

\begin{figure}[h]
	\centering
	\includegraphics[width=0.44\textwidth]{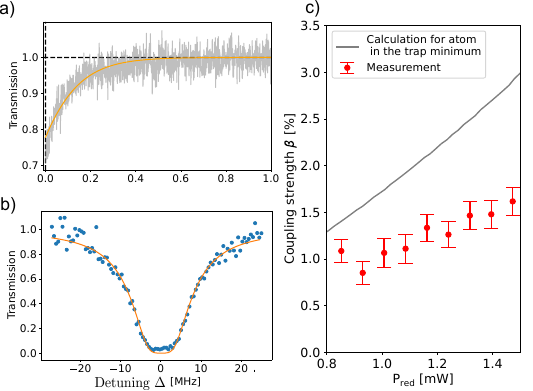}
	\caption{Measurement of the atom-light coupling strength $\beta$: a) We measure the total number of trapped atoms by monitoring the transmission of a laser  resonant to the $6S_{1/2} \; F=4 \rightarrow 6P_{3/2} \; F^{\prime}=3$ transition. We use a $1.0$-ms-long pulse through the nanofiber to depump the trapped atoms to the $6S_{1/2} \; F=3$ ground state. We deduce the number of trapped atoms $N = 165 \pm 13 $ from the time-dependent transmission as described in \cite{beguin_generation_2014}. b) Under the same conditions, we determine the on-resonance optical depth of OD $ = 6.9 \pm 0.8$ of the trapped ensemble via transmission spectroscopy as outlined in the main text. From this, we determine a coupling strength $\beta = 1.1\% \pm 0.2\%$ for the trap configuration with $P_\mathrm{red}=\unit[1]{mW}$ c) We measure $\beta$ for different values of $P_\mathrm{red}$.}
    \label{fig:Depumping}
\end{figure}

\begin{figure}[t]
\centering
\includegraphics[width=0.5\textwidth]{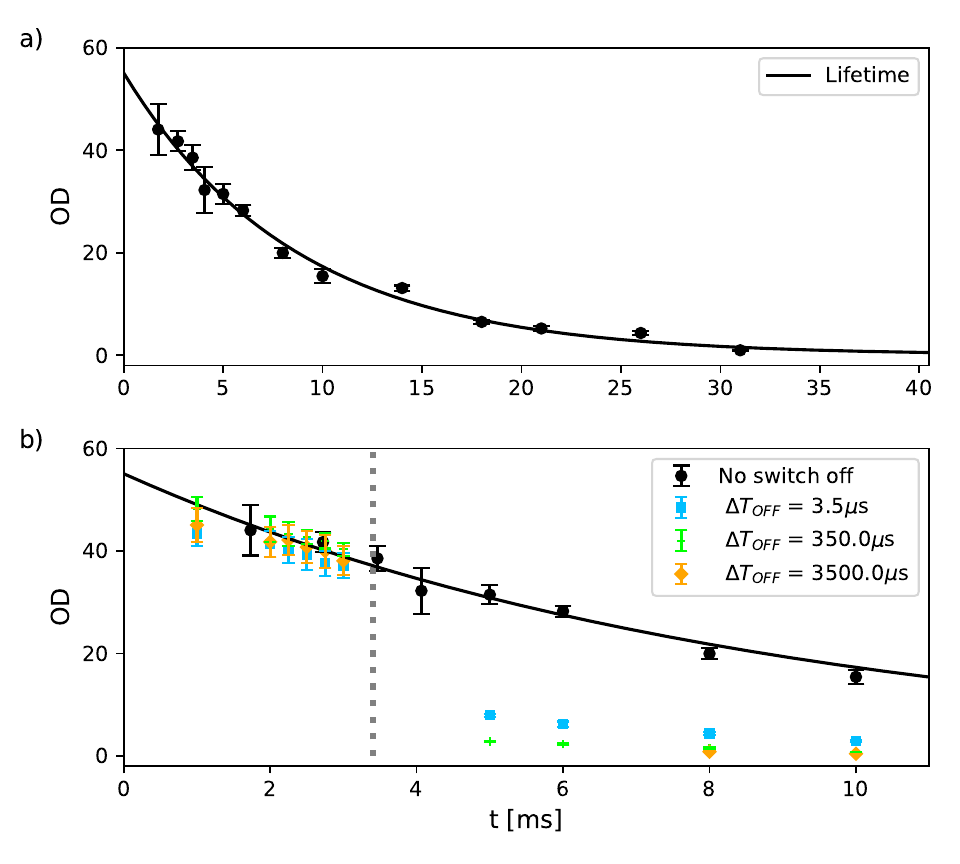}
\caption{Measurement of the OD of the ensemble of trapped atoms for different hold times after switching of the cooling laser: a) The exponential fit to the data yields a lifetime of $\tau = \unit[8.5]{ms} \pm \unit[0.4]{ms}$. b) The red-detuned trapping light field is switched off at $t=\unit[3.4]{ms}$ (dotted gray line). After $\Delta T_\mathrm{off}$ of \unit[3.5]{\textmu s} (cyan),  \unit[350]{\textmu s} (green), or \unit[3500]{\textmu s} (orange) the red-detuned trapping field is switched on again. A significant drop of the OD (colored) compared to the regular lifetime (black) is visible, confirming that most of the interrogated atoms during the lifetime measurement are indeed trapped. Experimental parameters are as in Fig.~\ref{fig:blue_potential}. The error bars indicate the $1\sigma$ standard deviation of the measured OD.}
\label{fig:blue_lifetime}
\end{figure}

\begin{figure}[h]
	\centering
	\includegraphics[width=0.5\textwidth]{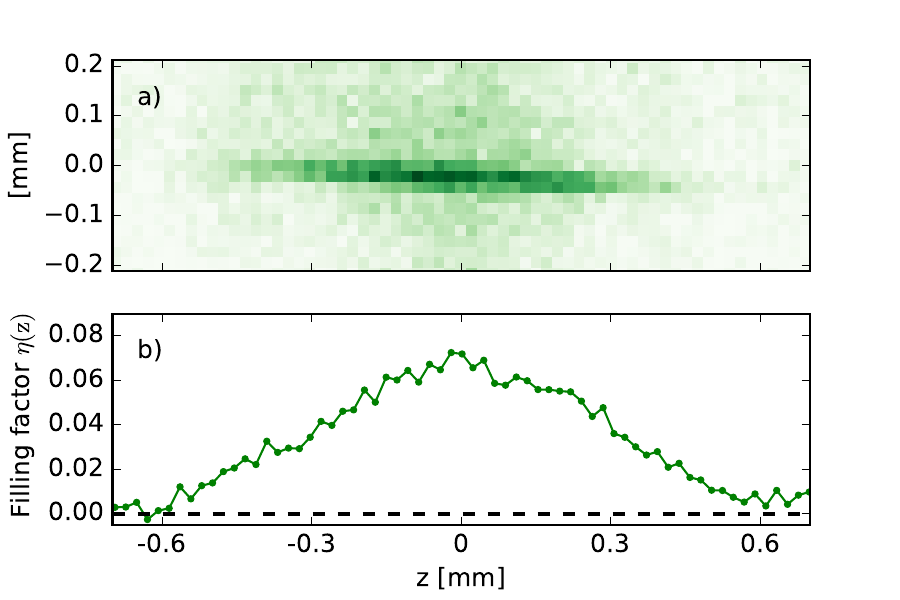}
	\caption{Measurement of the local filling factor $\eta (z)$. a) Fluorescence image of the trapped atoms along the nanofiber. The z-direction is aligned approximately with the nanofiber axis. b) Distribution of the local filling factor $\eta (z)$. We observe a Bell-shaped distribution of the trapped atoms with a maximum filling factor of $7\%$ in the center of the atomic distribution with a total atom number $N_{at} = 344 \pm 19$.}
	\label{fig:FillingFactor}
\end{figure}

\begin{figure}[t]
\centering
\includegraphics[width=0.5\textwidth]{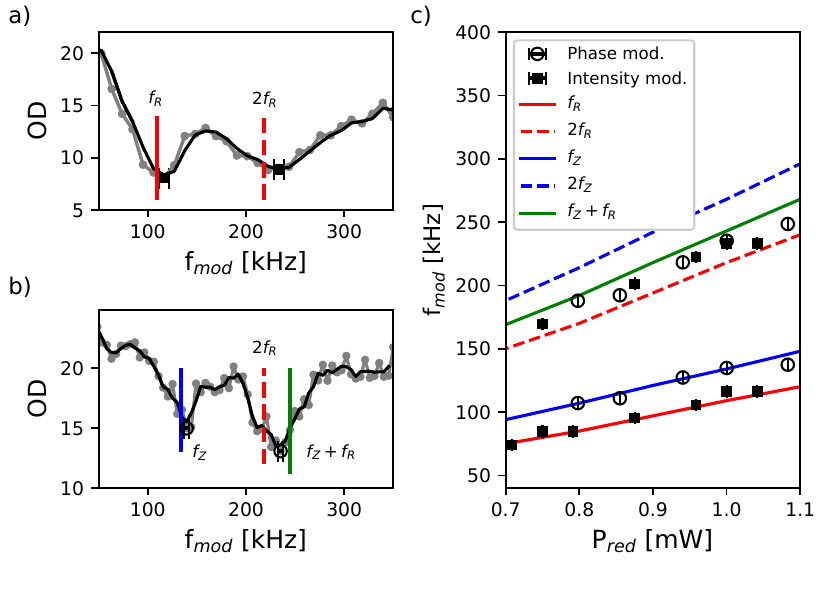}
\caption{Measurement of the trap frequencies: Measured OD when modulating a) the intensity of both blue-detuned trapping fields and b) their relative phase. The OD is determined by a fit of the transmission spectra after heating the atoms in the trap configuration as shown in Fig.~\ref{fig:blue_potential} ($P_\mathrm{red}=\unit[1]{mW}$). We identify the resonance frequencies by performing a moving average on the raw data (gray) resulting in the black curves, and then by searching for local minima. The determined local minima for the intensity modulation (black squares) and phase modulation (black circles) can then be compared to the calculated fundamental trap frequencies (vertical solid lines) and the second harmonics of the trap frequencies (vertical dashed lines) for the radial ($f_r$) and axial ($f_z$) motion. The step size of the applied modulation frequency for the intensity modulation is $\unit[10.2]{kHz}$, which we here take as an estimate of the uncertainty of the measured trap frequencies, yielding an error bar of $\pm \unit[5.1]{kHz}$. Likewise, the error bar for the phase modulation is estimated to be $\pm \unit[2.5]{kHz}$. c) Trap frequencies obtained from measurements identical to those shown in panels a) and b) for different powers of the red-detuned laser. We observe good agreement of corresponding theory predictions (solid and dashed lines) with the determined resonance frequencies for the intensity modulation (black squares) and for the phase modulation (black circles).
}
\label{fig:blue_frequencies} 
\end{figure}

The trapping scheme is based on a two-color nanofiber-guided dipole trap~\cite{LeKien_2004_Trap} realized by launching laser fields, which are red- and blue-detuned to the D2-transition of cesium, through a tapered optical fiber with a nanofiber-waist of $a =\unit[200]{nm}$ radius. The blue and red-detuned laser fields are launched from both ends into the nanofiber (see Fig.~\ref{fig:blue_trap}) and are tuned to $\lambda_\mathrm{blue} = \unit[685.4]{nm}$ and $\lambda_\mathrm{red} = \unit[935.7]{nm}$ close to the magic wavelengths of the D2-transition of cesium at $\lambda = \unit[852.3]{nm}$, so as to suppress the differential scalar light shifts between the ground and excited states~\cite{goban_demonstration_2012-2}. All fields are quasi-linearly polarized, where the transverse component of the electric field is aligned along the x-direction, as depicted in Fig.~\ref{fig:blue_potential} c). The interference of the two power-unbalanced counter-propagating blue-detuned trapping fields gives rise to a partial standing wave in the evanescent field, resulting in a dipole potential in the axial $z$-direction. Additionally, we launch two power-balanced counter-propagating red-detuned laser fields through the nanofiber to create an attractive dipole potential, thereby confining the atoms in all directions, radially, axially, and azimuthally. The two counter-propagating red-detuned laser fields are detuned from each other by $\sim \unit[150]{GHz}$ to prevent them from forming a standing wave pattern while at the same time minimizing the inhomogeneous Zeeman broadening due to the vector light shift~\cite{goban_demonstration_2012-2}. We note that the vector polarizability of cesium for the blue-detuned magic wavelength is two orders of magnitude smaller than that for the red-detuned magic wavelength~\cite{le_kien_dynamical_2013}. Therefore, the vector light shift for the blue-detuned dipole trap is negligible. Figure~\ref{fig:blue_potential} a) shows the total trapping potential, $U_{\mathrm{tot}}$, which is the sum of the blue and red-detuned dipole potentials as well as the van der Waals potential induced by the fiber surface~\cite{LeKien_2004_Trap}.
In order to provide a potential barrier that prevents atoms from escaping towards the fiber surface in the nodes of the blue-detuned standing wave, we use unbalanced powers of the two counter-propagating blue-detuned guided fields of $P_\mathrm{blue,1}=\unit[0.6]{mW}$ and of $ P_\mathrm{blue,2}=\unit[16]{mW}$, respectively, while the total power of the red-detuned light fields is $ P_\mathrm{red} = \unit[1]{mW}$. As shown in Fig.~\ref{fig:blue_potential} b) and c), this configuration creates two diametral one-dimensional arrays of trapping sites along the $z$-direction, where adjacent trapping sites are separated by $d = \lambda_\mathrm{blue} / (2 n) =\unit[300]{nm}$, which is $0.35 \lambda $. Here, $n = 1.14$ is the optical mode's effective refractive index. In the radial direction, the atoms are located around the local potential minimum at a distance of about \unit[350]{nm} from the fiber surface, see Fig.~\ref{fig:blue_potential} a). Finally, in the azimuthal $\varphi$-direction, the quasi-linear polarization of both red- and blue-detuned trapping fields leads to a modulation of the evanescent field intensity around the nanofiber, which results in the confinement of the atoms in the azimuthal direction \cite{kien_field_2004,vetsch_optical_2010}. For our trap configuration and parameters, the potential depth of our trap is approximately \unit{110}{\textmu K}. Due to the small trapping volume, our traps operate in the collisional blockade regime such that each trapping site contains at most one atom~\cite{schlosser_collisional_2002-1, Vetsch2012}.
\\
\indent
In order to load the atoms into the two arrays of trapping sites, we first prepare a cloud of cold cesium atoms by using a magneto-optical trap. This cloud overlaps with a section of the nanofiber, and in a subsequent molasses cooling stage the atoms are loaded into the nanofiber-based dipole trap. In order to estimate the number of loaded atoms, we measure the optical depth, (OD), of the trapped ensemble. To do this, we switch off the cooling laser and interrogate the atomic ensemble by launching a probe laser field through the nanofiber. We scan the laser frequency across the $6S_{1/2} \; F=4 \rightarrow 6P_{3/2} \; F^{\prime}=5$ D2-transition of the cesium. We then determine the OD of the ensemble by fitting a saturated Lorentzian function to the transmission data, as depicted in the inset of Fig. \ref{fig:blue_trap}. For this example, we obtain an optical depth of $\mathrm{OD} = 44.0 \pm 0.6$. We note that the probe power is small enough to neglect saturation effects. The optical depth relates to the number of trapped atoms, $N$,  via $\mathrm{OD} = 4 \beta N$. Here, $\beta = \Gamma_\textsc{wg}/\Gamma$ is the atom-light coupling constant, which describes the ratio of the spontaneous emission rate of an excited atom into the forward direction of the waveguide, $\Gamma_\textsc{wg}$, and the total emission rate, $\Gamma$. We determine the atom-light coupling constant $\beta$ by independently measuring the optical density, OD, and the number of trapped atoms, $N$. The latter is measured by using a depumping technique~\cite{beguin_generation_2014}. After loading, we launch a $1$-ms long pulse through the nanofiber that is resonant to $6S_{1/2} \; F=4 \rightarrow 6P_{3/2} \; F^{\prime}=3$ transition. As depicted in Fig.~\ref{fig:Depumping} a), we record the transmission as the atoms are pumped into the $6S_{1/2} \; F=3$ hyperfine ground state. From the time-dependent transmission, we deduce the trapped atom number $N = 165 \pm 13$. As shown Fig.~\ref{fig:Depumping} b), we then use the aforementioned transmission spectroscopy on the $6S_{1/2} \; F=4 \rightarrow 6P_{3/2} \; F^{\prime}=5$ transition to determine an OD of $6.9 \pm 0.8$ of the atomic ensemble under the same experimental conditions. From this, we infer a coupling strength of $\beta = 1.1\, \% \pm 0.2 \, \%$ for the atoms in our trap configuration. We repeat the measurement for different powers of the red-detuned trapping field. As depicted in Fig.~\ref{fig:Depumping} c), we measure a coupling strength $\beta$ between $0.85\, \%$ and $1.85\, \%$, depending on the total power of the red-detuned trapping light fields. We attribute this dependence to the varying distance of the trap minima from the nanofiber surface. The grey line in Fig.~\ref{fig:Depumping} c) shows the calculation of the coupling strength $\beta$ for a ground state atom. A finite temperature of the atoms leads to a larger mean distance of the atoms from the fiber surface in the trap potential, which is asymmetric around its minimum (see Fig.~\ref{fig:blue_potential} a)), making the smaller observed $\beta$-value plausible.
\\
\indent
The lifetime of the atoms is determined by measuring the OD as a function of the delay time, $t$, between the loading and the center of the $\unit[1]{ms}$ probe pulse. We obtain the data shown in Fig.~\ref{fig:blue_lifetime} a). Fitting an exponential decay of the OD, we find a trap lifetime of $\tau = \unit[8.5]{ms} \pm \unit[0.4]{ms}$. In order to confirm that the measured transmission signal is indeed due to trapped atoms rather than untrapped atoms from the residual atomic cloud surrounding the nanofiber, we carry out an additional measurement. Its experimental sequence is similar to the one used for the above lifetime measurement, with the difference that at $t = \unit[3.4]{ms}$, see gray dashed line in Fig.~\ref{fig:blue_lifetime} b), we switch off the red-detuned trapping field for a duration $\mathrm{\Delta T_{off}}$. After a duration $\mathrm{\Delta T_{off}}$ of \unit[3.5]{\textmu s} (cyan), \unit[350]{\textmu s} (green) or \unit[3500]{\textmu s} (orange), we switch the red-detuned trapping field on and measure the remaining OD. As shown in Fig.~\ref{fig:blue_lifetime} b), we observe a sharp drop of the OD when the red-detuned laser is switched off. This indicates that the atoms were indeed trapped before the switch-off because the latter does not affect the density of the residual atomic cloud surrounding the nanofiber. For the shortest switch-off time, we observe a higher residual OD than in the other two cases. This can be explained by the fact that $\mathrm{\Delta T_{off}} = $ \unit[3.5]{\textmu s} is comparable with the trap oscillation period, see below, such that some atoms are recaptured in the dipole trap when switching the trapping field back on.
\\
\indent
In the selective radiance regime, we expect a narrowing of the transmission spectrum. However, we do not observe such a narrowing in the transmission spectroscopy, see inset of Fig.~\ref{fig:blue_trap} and Fig.~\ref{fig:Depumping}. To estimate whether this is due to the too low local filling factor, $\eta(z)$, we measure the extent of the trapped atomic ensemble by imaging it along the fiber depicted in Fig.~\ref{fig:FillingFactor} a). For this purpose, we probe the atoms for 1 ms with a weak fiber-guided pulse that is detuned by $3 \Gamma$ from the $6S_{1/2} \; F=4 \rightarrow 6P_{3/2} \; F^{\prime}=5$ D2-transition. The probe parameters are chosen in such a way that we do not alter the trap potential and excite the atoms uniformly. We image the fluorescence of the atoms using an electron-multiplying charge-coupled device (EMCCD) camera and determine their density distribution, where the width of a logical pixel corresponds to \unit[22]{\textmu m} in the object plane. Afterwards, we perform the aforementioned depumping measurement to determine the number of trapped atoms. From this, we calculate a maximum filling factor of $7\,\%$ in the center of the distribution of the trapped atoms as depicted in Fig.~\ref{fig:FillingFactor} b). To avoid overexposure of the camera by the molasses light, we image the atoms with a delay at which the OD is about 2.5 times smaller than directly after loading. However, even when considering the OD at zero delay, the maximum filling in the center of the trapped atomic ensemble remains smaller than $20\, \%$, well below the limit of the collisional blockade regime~\cite{schlosser_collisional_2002-1, Vetsch2012}. Furthermore, we expect the regions of lower atomic line density surrouding the center of the trapped atomic ensemble to suppress signatures of selective radiance.
\\
\indent
To finalize the study of this new trap configuration, we now characterize the confinement of the atoms in the trap by measuring the trap frequencies along the radial ($f_r$) and axial ($f_z$) direction. To do this, we perform loss spectroscopy based on a resonant heating technique~\cite{savard_laser-noise-induced_1997} using an acoustic-optic modulator (AOM) and an electro-optical modulator (EOM) as depicted in Fig.~\ref{fig:blue_trap}. First, we modulate the intensity of the blue-detuned trapping field with the AOM before it is split and launched in the two counter-propagating directions through the nanofiber. The modulation of the intensity translates into a modulation of the center position of the trapping potential in the radial direction. After loading the trap, we modulate the potential for \unit[2]{ms} and then measure the OD. When the modulation frequency $f_\text{mod}$ reaches a harmonic of the trap frequency, atoms are heated out of the trap and the OD decreases. 
To identify the resonance frequencies for which the atoms are subject to maximal heating, we first perform a moving average on the OD extracted from the transmission data to mitigate the effect of shot-to-shot fluctuations of the number of trapped atoms. Afterwards, we apply a search-for-local-minima algorithm on the smoothed data set. The results of the measurement are shown in Fig.~\ref{fig:blue_frequencies} a). We observe local minima of the OD at $\unit[116]{kHz}$ and $\unit[232]{kHz}$ (black squares). The measured resonances agree well with the theoretically predicted fundamental and second harmonic radial trap frequencies at $f_r = \unit[109]{kHz}$ and $2 f_r = \unit[218]{kHz}$ (red vertical lines). Notably, we can effectively excite the atomic motion at twice the radial trap frequency due to the trapping potential's strong anharmonicity, as shown in Fig.~\ref{fig:blue_potential}a), and also caused by an inherent modulation of the radial trap frequency, giving rise to parametric heating. In order to measure the axial trap frequency, we use an EOM that modulates the phase of one of the two blue-detuned trapping fields, see Fig.~\ref{fig:blue_trap}. The EOM is aligned such that it only modulates the phase of the corresponding blue-detuned trapping field without affecting its polarization. This phase modulation, applied for a duration of \unit[5]{ms}, causes a sinusoidal modulation of the position of the standing wave potential along the $z$-direction. Figure~\ref{fig:blue_frequencies} b) shows the measured OD as a function of the modulation frequency. We observe two local minima at $\unit[139]{kHz}$ and at $\unit[235]{kHz}$ (black circles). The former is close to the theoretically predicted fundamental axial trap frequency of $f_z = \unit[134]{kHz}$ (blue vertical line), while the latter is close to the the second harmonic of the radial frequency at $2f_r = \unit[218]{kHz}$ (red dashed vertical line) as well as to the sum frequency $f_r + f_z = \unit[245]{kHz}$ of the axial and radial motion (green vertical line). This resonance at the sum frequency is, indeed, expected due to the non-separability of the nanofiber-based trapping potential. This implies that the axial and radial motions are coupled~\cite{meng_near-ground-state_2018}, so that an oscillation along the axial direction also leads to an oscillation in the radial direction. We attribute the observed width of the resonances to the anharmonicity of the trap potential and the resulting energy-dependent trap frequency. In order to perform a systematic characterization of the new trap configuration, we repeat the measurement of the trap frequencies for different powers of the red-detuned trapping field. The resulting trap frequencies are shown in Fig.~\ref{fig:blue_frequencies} c). The experimental observations agree well with the theoretical predictions over the entire parameter range of $P_\mathrm{red}$. In this work, we focus on the radial and axial degrees of freedom as an accurate measurement of the azimuthal trap frequency is challenging. The potential is quite shallow along the azimuthal coordinate with a trapping frequency between $\unit[10]{kHz}$ and $\unit[20]{kHz}$ for our range of red-detuned trap fields. Furthermore, both implemented drive mechanisms excite the azimuthal degree of freedom only weakly. 
\newline
\\
In summary, we have experimentally demonstrated magic-wavelength trapping of laser-cooled atoms in a novel type of nanofiber-based two-color dipole trap. By using a blue-detuned partial standing wave and two red-detuned running waves, we create two diametral one-dimensional arrays of trapping sites for cesium atoms. In this trapping potential, the spacing of two adjacent sites is $d \simeq 0.35 \lambda$, i.e.~only 35~\% of the resonant free-space wavelength of the cesium D2 line. We determine the lifetime of the atoms in this new trapping configuration to be $\tau = \unit[8.5]{ms} \pm \unit[0.4]{ms}$ and perform a systematic characterization of the radial and axial trap frequencies which agree well with the expectations. 
\\
\indent
This realization and characterization of a sub-$\lambda/2$-spaced standing wave nanofiber-guided dipole trap is an important step towards the experimental observation of novel collective radiative effects, such as selective radiance in 1D atomic arrays. In a next step, the filling factor of the trapping site has to be increased, from currently about $20\,\%$ to ideally close to unity. This can be achieved, by increasing the loading rate using, e.g., more efficient cooling mechanisms, such as $\Lambda$-enhanced gray molasses cooling~\cite{PhysRevA.98.033419}. Alternatively, even with a lower filling factor, signatures of selective radiance are predicted and should for example, show up as a significantly increased reflection of light from the ensemble. In the standard two-color nanofiber-based trap, the power reflection coefficient of a guided light field is on the order of $\beta^2\sim10^{-4}$. However, already a small, defect-free array of about $10$ adjacent atoms would give rise to a reflection coefficient of close to 10~\%, corresponding to an increase by three orders of magnitude~\cite{ChangHummerPrivate}. Such arrays are highly likely to be randomly present as part of large nanofiber-trapped ensembles with moderate filling factor. By removing all atoms except those in the highest density region from the trap, signatures of selective radiance could be observed.

\section*{Acknowledgment}
M. C. and M. S. acknowledge support from the European Commission (Marie Skłodowska-Curie IF Grants Nos.~101029304 and IF Grant No.~896957). H. L. acknowledges funding by the European Commission (ERC Grant No.~101071882). We acknowledge funding by the Alexander von Humboldt Foundation in the framework of the Alexander von Humboldt Professorship endowed by the Federal Ministry of Education and Research, funding by the European Commission under the projects DAALI (No.~899275), and SuperWave (ERC Grant No. 101071882), as well as funding by the Einstein Foundation
(Einstein Research Unit on Quantum Devices). We thank Darrick Chang and Daniel Hümmer for insightful discussion and calculations of selective radiance in our nanofiber-platform.  We thank Christoph Clausen, Christian Liedl, and Sebastian Pucher for developing and improving the trap potential calculations. 

\bibliography{BlueTrap.bib}

\end{document}